\documentclass[conference]{IEEEtran}
\usepackage{amsmath,amsfonts,amssymb,amsthm}
\usepackage{times}
\usepackage{comment}
\usepackage{xcolor}
\usepackage{graphicx}
\usepackage{subfigure}
\usepackage{multirow}
\usepackage[numbers]{natbib}
\usepackage{multicol}
\usepackage[bookmarks=true]{hyperref}
\usepackage{makecell} 

\usepackage{floatrow}
\usepackage[linesnumbered,ruled]{algorithm2e}
\usepackage{txfonts}

\begin{document}

\title{Coordinating tiny limbs and long bodies: geometric mechanics of diverse undulatory lizard locomotion 
}

\author{\authorblockN{Baxi Chong\authorrefmark{1},
Tianyu Wang\authorrefmark{1},
Eva Erickson\authorrefmark{1}, 
Philip J Bergmann\authorrefmark{2}, 
Daniel I. Goldman\authorrefmark{1}}\\
\authorblockA{\authorrefmark{1}School of Physics, Georgia Institute of Technology,\authorrefmark{2}Department of Biology, Clark University}}

\maketitle

\begin{abstract}

Although typically possessing four limbs and short bodies, lizards have evolved diverse body plans including but not limited to elongate trunks with tiny limbs. These elongate morphologies are hypothesized to aid locomotion in cluttered and fossorial environments. However, mechanisms of propulsion in such forms – e.g. the use of body and/or limbs to interact with the substrate – and potential body/limb coordination remain unstudied. Here, we use biological experiments, a geometric theory of locomotion, and robophysical models to comparatively investigate body-limb coordination in a diverse sample of lizard morphologies. Locomotor field studies in short limbed, elongate lizards (\textit{Brachymeles} and \textit{Lerista}) and laboratory studies of fully limbed lizards (\emph{Uma scoparia} and \textit{Sceloporus olivaceus}) and a snake (\textit{Chionactis occipitalis}) reveal that body wave dynamics can be described by a combination of standing and traveling waves; the ratio of the amplitudes of these components is inversely related to the degree of limb reduction and body elongation. The geometric theory helps explain our observations, predicting that the advantage of traveling wave body undulations (compared with a standing wave) emerges when the dominant thrust generation mechanism arises from the body rather than the limbs. We test our hypothesis in biological experiments by inducing use of traveling waves in stereotyped lizards via modulating the ground penetration resistance. Study of a limbed/undulatory robophysical model demonstrates that a traveling wave is beneficial when thrust is generated by body-environment interaction. Our models could be valuable in understanding functional constraints on the evolutionary process of elongation and limb reduction in lizards, as well as advancing robot designs. 

\end{abstract}

\IEEEpeerreviewmaketitle
\section{Introduction}

Recent studies demonstrated that the evolution of body elongation and limb reduction is convergent among most major lineages of vertebrates, including not but limited to fish \cite{ward2007evolution}, amphibians \cite{parra2001extreme}, reptiles \cite{brandley2008rates}, and even mammals \cite{buchholtz2007vertebral}. Of particular interest, in Squamate reptiles (lizards and snakes) snake-like body shapes were reported to have independently evolved at least 25 times \cite{wiens2006does,sites2011phylogenetic}. While the exact selective pressures for this evolutionary transition remain a mystery, prior studies revealed possible advantages of certain body plans in navigating their corresponding environments \cite{gans1975tetrapod,wiens2006does,bergmann2020locomotion,morinaga2020evolution}. One of the most popular hypotheses is that limbless forms have evolved as an adaptation for fossoriality (underground environment) or an adaptation for cluttered environments, \cite{rieppel1988review,simoes2015visual}, as snakes are believed to have evolved from a fossorial ancestor \cite{wiens2006does,rieppel1988review}.

\begin{figure}[t]
\centering
\includegraphics[width=0.9\linewidth,trim=0 0 0 0,clip]{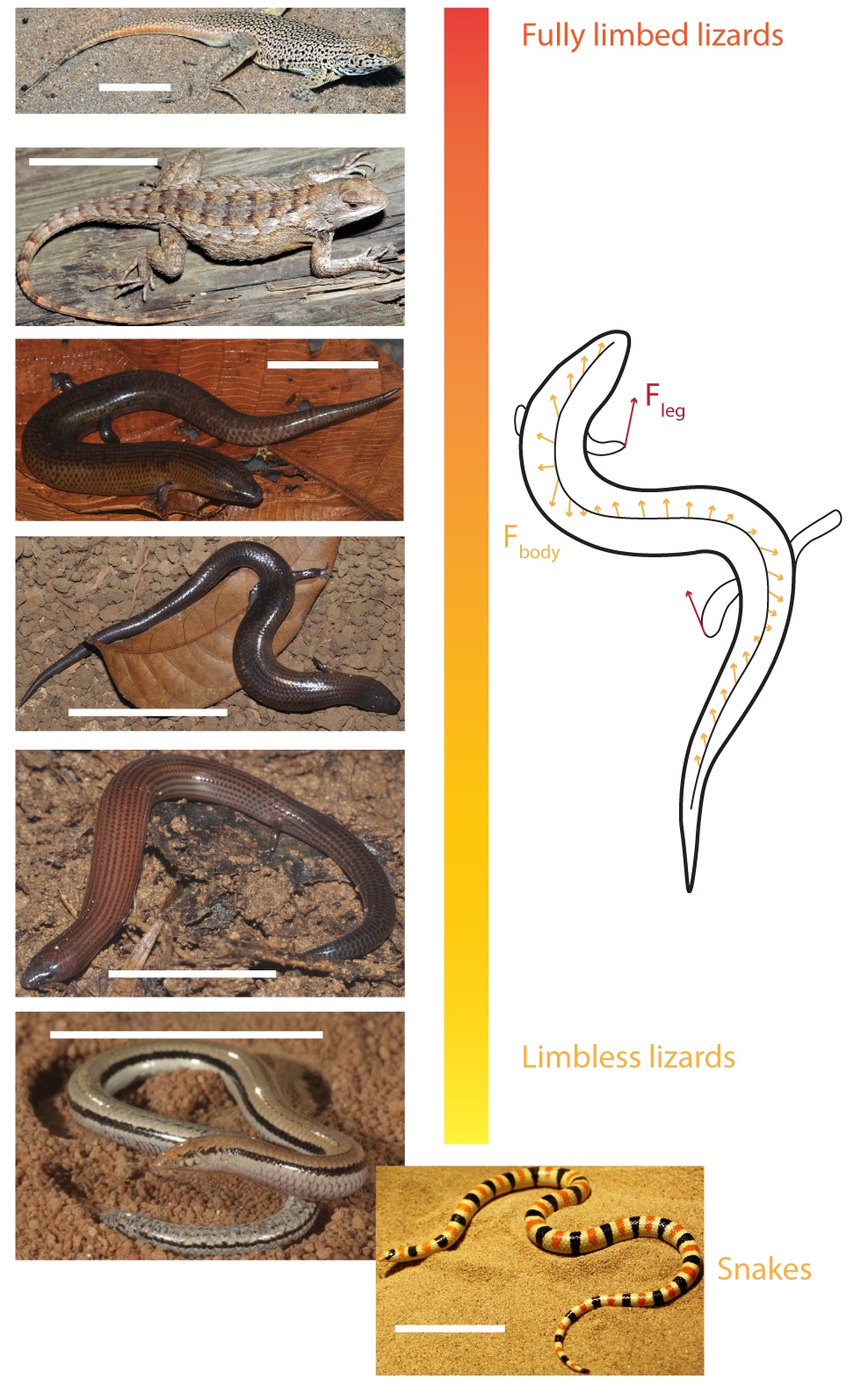}
\caption{\textbf{Target and model systems for understanding the role of body undulation in the lizard body elongation and limb reduction continuum} (Left) Extant short limbed, elongate lizards (\emph{B. muntingkamay}, \textit{B. kadwa}, \textit{B. taylori}) in comparison with fully limbed lizards (\textit{U. scoparia} and \textit{S. olivaceus}) and limbless/almost limbless species (\textit{C. occipitalis} and \textit{L. praepedita}). Scale bars indicate 5 cm. (Right) An illustrative diagram of the thrust generation in short limbed, elongate lizards: the thrust generated by limb retraction is labeled in red arrows, the thrust generated by body undulation is labeled in yellow arrows.}
\label{fig:allfigs}
\end{figure}

Transitions in body morphology are one of many aspects of evolutionary adaptation for locomotion. Another crucial but less studied aspect in such adaptation is how to most effectively control and coordinate movements of body morphologies during locomotion. In other words, how do the transitions in body morphology correspond to transitions in the locomotion pattern? For example, the use of traveling waves has been observed in Squamate reptiles across different morphology. Specifically, snakes (and presumably limbless lizards) primarily use traveling wave body undulations to generate thrust \cite{jayne1988muscular}. While fully limbed lizards use standing wave body bending \cite{farley1997mechanics} to assist limb retraction at low speed, fully limbed lizards and salamanders use traveling waves as their speed increases~\cite{roos1964lateral,daan1968lateral,edwards1977evolution,ritter1992lateral,frolich1992kinematic}. It is hypothesized that a traveling wave can be beneficial at high locomotor speeds because it could transmit force along the axis of progression~\cite{ritter1992lateral}. Finally, in species with shorter limbs, the role of body bending is believed to directly provide propulsive force~\cite{ritter1992lateral}. Yet, it is still unknown the relative advantage between standing waves and traveling waves in those lizards with short limbs.
Here, we focus on the last case, which we refer to as terrestrial swimming, in the context of the body elongation and limb reduction continuum.
We are among the first to consider the degree of standing/traveling wave as a continuum, and to test how it relates these intermediate body forms. A number of important question remain unanswered:
how should lizards with short limbs relative to body length (short limbed, elongate lizards) coordinate their body movements with their limbs? How do body movements relate to the degree of body elongation and limb reduction? Answering these questions will not only establish a relationship between what they have (the body morphology) and how they move (the body-limb coordination)~\cite{nyakatura2019reverse,nyakatura2014bridging,rieser2021functional,mcinroe2016tail}, but also facilitate our understanding of the locomotor implications of the evolution of snake-like forms~\cite{bergmann2019convergent,morinaga2020evolution}.

\begin{figure*}
\centering
\includegraphics[width=1\linewidth,trim=0 0 0 0 ,clip]{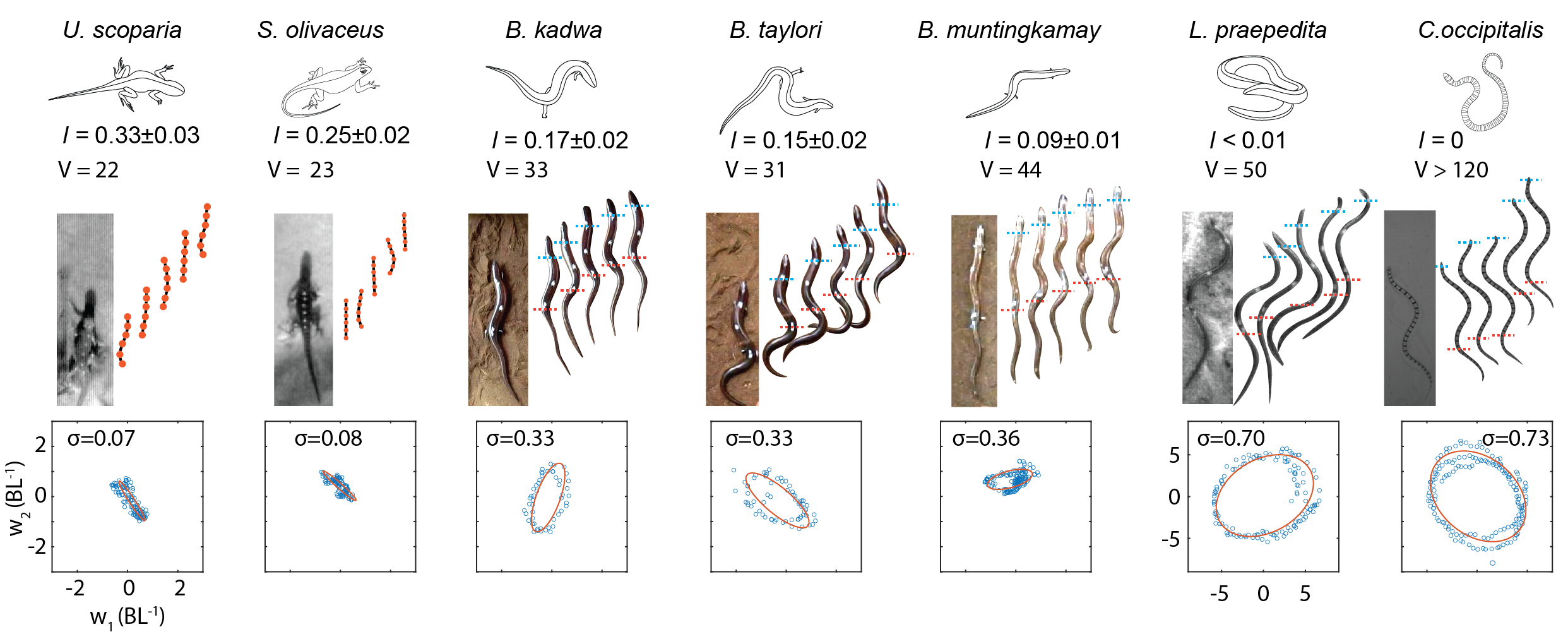}
\caption{\textbf{The diversity of body waves in the body elongation and limb reduction continuum}. (top) Photos of species and the snapshots of their body motion during one period (at a scale of seconds) of locomotion. Seven species were studied (from left to right): \textit{U. scoparia}, \textit{S. olivaceus}, \textit{B. kadwa}, \textit{B. taylori}, \textit{B. muntingkamay}, \textit{L. praepedita}, and \textit{C. occipitalis}. The relative limb size (\textit{l}: the hind limb length normalized by SVL) and number of presacral vertebrae ($V$) for each species are labeled \cite{bergmann2012vertebral,bergmann2019convergent}. (bottom) The projections of body curvature into the reduced shape space and the estimation of $\sigma$ for each animal. Units of axes are identical to the left panel.}
\label{fig:animal}
\end{figure*}

The challenges of studying short limbed, elongate lizard locomotion lie in  discerning the coordination between body undulation and limb retraction.
During terrestrial locomotion, locomotors generate thrust by pushing on the substrate. There are many mechanisms for generating thrust, such as lateral body undulation (commonly seen in limbless locomotors) and limb retraction (commonly seen in limbed locomotors). In most cases, a locomotor relies on a single mechanism to generate thrust. 
However, in lizards with short limbs, limbs which cannot support the animal's body weight, the two mechanisms (limb retraction and body undulation) can coexist, requiring proper coordination of both propulsive mechanisms.
Therefore, the support of body weight should be properly distributed between the ventral surface of the body and the limbs to facilitate effective thrust-generation mechanics. 
In this way, we believe the key to studying short limbed, elongate lizard locomotion is to understand the coordination of two thrust-generating mechanisms.




To address these questions, we take a comparative biological, robophysical, and theoretical modeling approach. We compile a collection of high-speed videos of a spectrum of lizard body forms collected in both field and laboratory settings. Through the use of neural network markerless tracking~\cite{nath2019using}, we analyze the data and reveal a striking diversity in body undulation dynamics. Specifically, we find that body undulation in lizards with short limbs is a linear combination of a standing wave and a traveling wave; and that the ratio of the amplitudes of these two components is inversely related to the degree of limb reduction and body elongation. 
The fact that our animals move in highly damped environments, where frictional forces dominate over inertial forces, allows the use of the geometric mechanics framework \cite{wilczek1989geometric,hatton2011geometric} to explain wave dynamics and body-limb coordination. 
This geometric mechanics theory, which replaces laborious calculation with diagrammatic analysis, rationalizes the advantage of using traveling waves in short limbed elongate lizards, and predicts that such advantages emerge when the primary thrust generation source shifts from the limbs to the body.
We test our hypothesis with biological and robophysical experiments by manipulating the substrate on which fully limbed lizards move, and with robophysical experiments by controlling the body and limb thrust mechanism.

\section*{Results}
\begin{figure}[h!]
\centering
\includegraphics[width=0.8\linewidth,trim=0 0 0 0 ,clip]{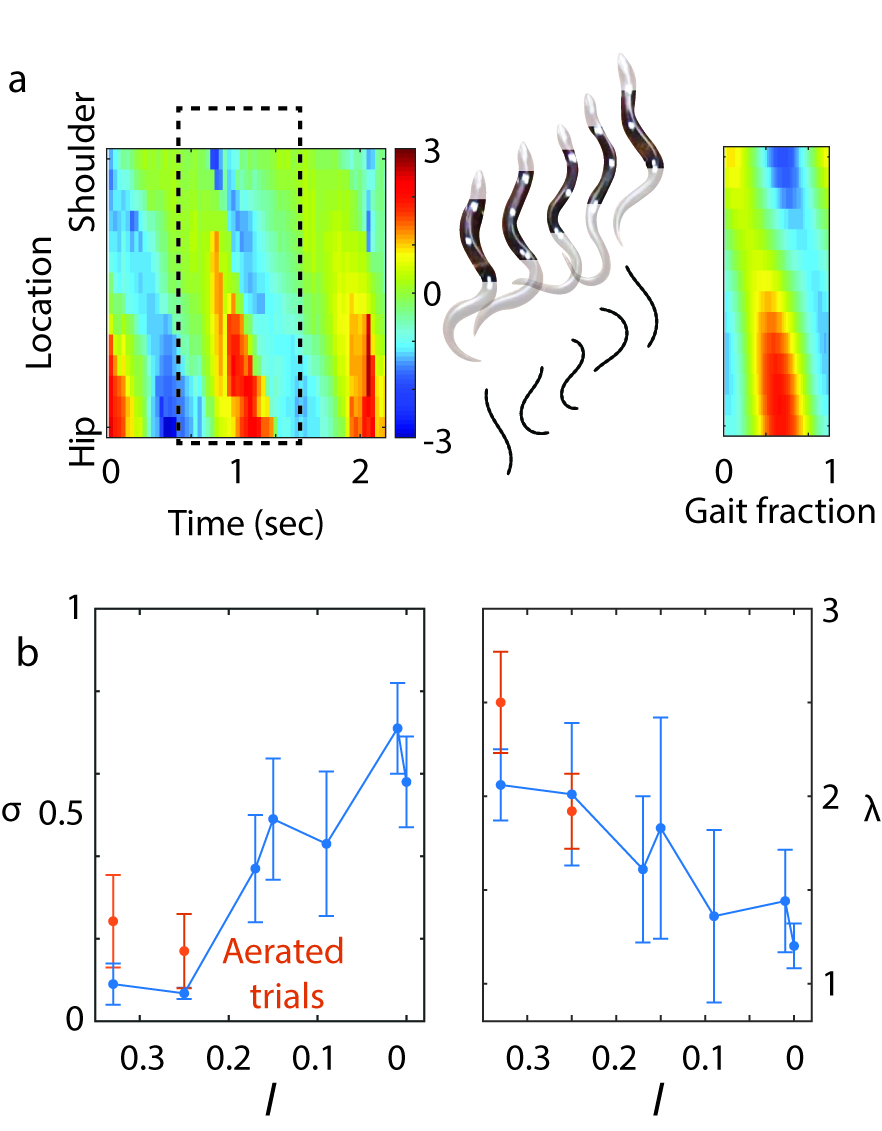}
\caption{\textbf{From standing wave to traveling wave} (a) Comparison between the original body curvature profile of \textit{B. taylori} and the reconstructed body curvature profile over a gait cycle from the estimated wavelength $\lambda$ and flatness $\sigma$. The units of the colorbar are SVL$^{-1}$. 
(b) The relationship between the locomotion parameters ($\sigma$ and $\lambda$) and morphology parameter (the relative hind limb length $l$). Red points with error bars correspond to the locomotion parameters of \textit{U. scoparia} and \textit{S. olivaceus} on an aerated granular medium to reduced the resistive force of the media.  Note that we use $l=0.01$ for \emph{L. praepedita} on the plot and that the abscissa is reversed (descending left to right) to correspond to Fig. 2.
}
\label{fig:analysis}
\end{figure}

\subsection*{Diversity in lizard body movements} 
We investigated three short limbed, elongate species with equally developed fore and hind limbs \cite{siler2011evidence,bergmann2019convergent} (\emph{Brachymeles kadwa}, \emph{Brachymeles taylori}, and \emph{Brachymeles muntingkamay}) and compared them with fully limbed lizards (\emph{Uma scoparia} and \textit{Sceloporus olivaceus}) and limbless species (the almost limbless lizard \emph{Lerista praepedita} and the shovel-nosed snake \textit{Chionactis occipitalis}). These species were chosen because they form a spectrum of limb reduction and body elongation (Fig. \ref{fig:allfigs}). The relative limb size is defined as the hind limb length normalized by SVL (snout vent length). The number of presacral vertebrae is a measure of elongation~\cite{bergmann2019convergent} (Fig. \ref{fig:animal}). We recorded field videos of these species moving on granular media (consisting of soil and poppy seeds), and compared the kinematics of their body movements. The snapshots of their body postures during locomotion are compared in Fig. \ref{fig:animal} (middle panel). Qualitatively, we observed the node\footnote{The point in the body which has zero body curvature} of body bending is almost stable in fully limbed lizards (at the shoulder and hip), and propagates from snout to cloaca in shovel-nosed snakes. Interestingly, in short limbed, elongate species, one of the nodes is almost stable near the snout, and the other node propagates from the mid-body to tail.

We considered locomotion as a properly coordinated sequence of ``self-deformation" (the sequence of internal shape changes) that generates thrust (self-propulsion\footnote{We will explain further the terminology of self-propulsion and self-deformation when we discuss geometric mechanics}) via interactions with substrates. Prior work \cite{chong2019hierarchical,chong2021coordination,hatton2013geometric,gong2016simplifying,rieser2019geometric} suggested that despite possessing high dimensionality, the essence of self-deformation can be described by a linear combination of shape basis functions. Consider the body curvature\footnote{Body curvature is the inverse of radius of curvature} $\kappa(s,t)$ at time $t$ and location $s$ ($s = 0$ denotes the snout in snakes (or the shoulder in lizards) and $s=1$ denotes the cloaca in snakes (or the hip in lizards)). Thus, the body curvature profile can be approximated by:

\begin{equation}
\kappa(s,t)=w_1 (t)\sin{(2\pi\xi s)} + w_2 (t)\cos{(2\pi\xi s)},
\end{equation}

\noindent where $\xi$ is the spatial frequency of body undulation obtained from direct fitting ($1/\xi$ denotes the wavelength, $\lambda$, in the unit of SVL); $w_1(t)$ and $w_2(t)$ are the reduced shape variables describing the instantaneous shape of the locomotor at time~$t$. In this way, we can map the original high-dimensional body curvature profile $\kappa(s,t)$ into a space spanned by $w_1$ and $w_2$. In pure standing waves, the body curvature trajectory in the reduced shape space can be described as a flattened ellipse (with eccentricity $e\rightarrow 1$). In pure traveling waves, the body curvature trajectory in the reduced shape space can be described as a circle (with eccentricity $e\rightarrow 0$). In this way, an elliptical trajectory can be considered a linear combination of the flattened ellipse path and the circular path, the ratio of which can be quantified by the flatness ($\sigma = \sqrt{1-e^2}$), where $\sigma = 0$ denotes a pure standing wave and $\sigma = 1$ denotes a pure traveling wave. We compared the gait trajectory for species ranging from fully limbed to limbless animals in Fig. \ref{fig:animal} (bottom panel), where we observed a transition from a flattened ellipse in stereotyped lizards to a circle in snakes.

To quantitatively measure the flatness of the gait trajectories in the reduced shape space, we fit these trajectories with oriented ellipses. To test the accuracy of the fitting, we compared the original body undulation profile (collected from tracking in field videos, left panel) and the fitted body undulation profile (from a reconstruction of the ellipses in reduced shape spaces, right panel) in Fig. \ref{fig:analysis}a. Interestingly, we observed that $\sigma$ increases and $\lambda$ decreases as the limb size decreases, indicating a transition from standing wave to traveling wave as the limb size decreases (and number of presacral vertebrae increases) (Fig. \ref{fig:analysis}b).

\subsection*{Wave dynamics are key to body-limb coordination}

\begin{figure}[!ht]
\centering
\includegraphics[width=1\linewidth]{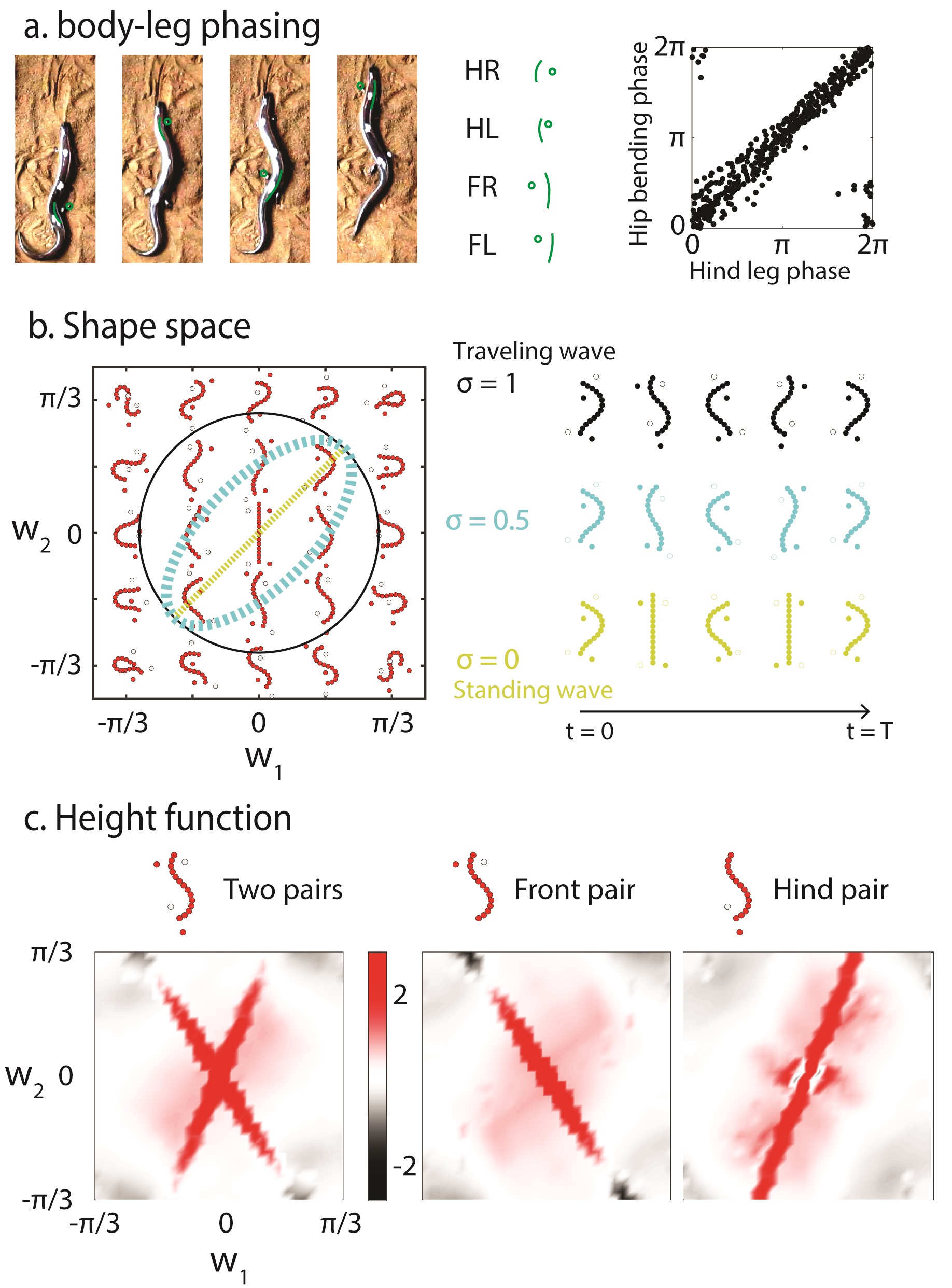}
\caption{\textbf{Geometric mechanics analysis of the body-limb coordination in short limbed, elongate lizards} (a) The limb movement in short limbed, elongate lizards follows the lateral couplet sequence (FR-HL-FL-HR). The phase relationship of hip bending and hind limb movements are plotted in the right side panels. (b) The shape space for short limbed, elongate lizards. The body movements are prescribed by the reduced shape variable $w_1$ and $w_2$, and the limb contact states are inferred from the body movements. Gaits can be represented by closed-loop paths in the shape space. A standing wave gait path, a traveling wave gait path, and an intermediate wave gait path are compared. (c) Height functions to investigate the body undulation in lizards with intermediate limbs. (left) Two strips emerged in the height function for short limbed, elongate lizards, such that a circular gait path can enclose significantly more surface than a flattened elliptic gait path. To further understand the two stripes, we calculated the height function for hypothetical lizards with one pair of limbs near the head (middle panel) and near the tail (right panel). Each stripe is associated with a pair of limbs, in which case a flattened elliptic gait path can enclose sufficient surface in the height function. The units of the colorbar are ($10^{-3}\times \text{SVL}^{-1}/\text{rad}^{2}$).}

\label{fig:coordination}
\end{figure}

We further analyzed the limb movement in the short limbed, elongate species (\textit{B. kadwa} and \textit{B. taylori}). Snapshots showing the body posture during the touchdown of each foot are illustrated in Fig. \ref{fig:coordination}a (left panel). The limb movements in short limbed, elongate species follow the sequence: FR-HL-FL-HR (F, H, R, and L represent fore, hind, right, and left respectively). Specifically, the hind leg leads the fore leg on the same side by $38.1\pm6.7\%$ of a period, in the category of lateral couplet sequence according to~\cite{hildebrand1965symmetrical}. Further, for each leg, the ground-contact (stance phase) duration is approximately the same as the ground-lifting (swing phase) duration, indicating the duty factor (the fraction of a period that each limb is on the ground) is approximately 0.5.

\begin{figure}[ht]
\centering
\includegraphics[width=1\linewidth,trim=0 0 0 0 ,clip]{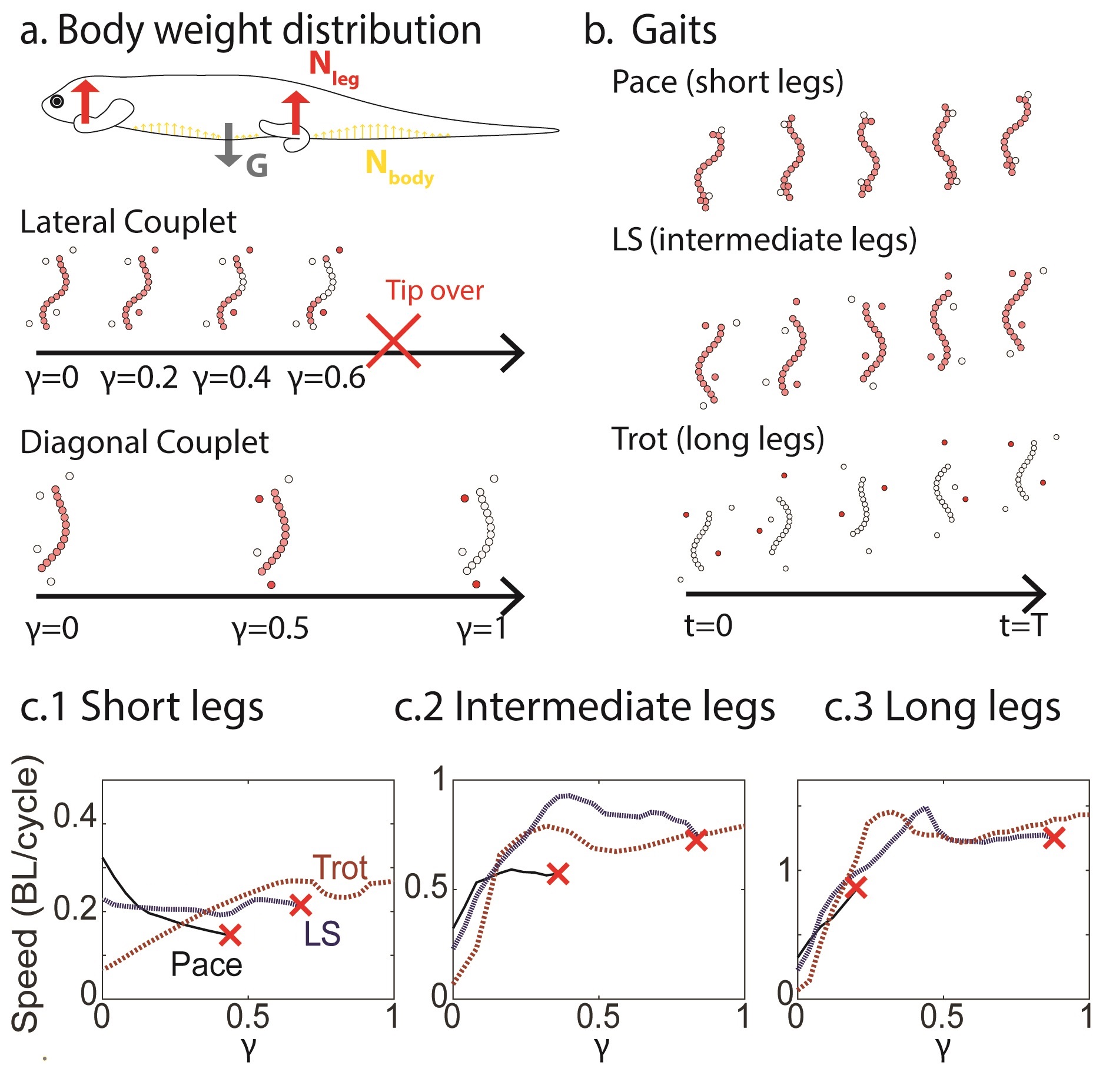}
\caption{\textbf{The weight distribution role of limbs in lizard locomotion} (a) The body weight can be supported by the limbs and the body; $\gamma$ indicates the fraction of body weight supported by limbs. (b) Three typical gaits: the pace gait (duty factor = 0.5, leg phase shift = 0) implemented by lizards with short limbs, the LS (lateral sequence: duty factor = 0.5, leg phase shift = 0.25) gait implemented by lizards with intermediate limbs, and the trot gait (duty factor = 0.5, leg phase shift = 0.5) implemented by lizards with long limbs. (c) The relationship between $\gamma$ and speed for (solid black curves) pace, (dashed blue curves) LS, and (dashed red curves) trot gaits on lizards with (c.1) short, (c.2) intermediate, and (c.3) long limbs. Potential tip-overs are indicated by a red cross.}
\label{fig:distribution}
\end{figure}

We also noticed that during a foot touchdown, the local body element develops maximal curvature (in convex direction towards the leg) to increase its reach (Fig. \ref{fig:coordination}a, middle panel), which is consistent with observations of other quadrupedal locomotors \cite{Ijspeert2001,chong2021coordination}. This observation indicates that the fore (hind) limb movement should be in phase with shoulder (hip) bending. We quantify this observation by showing the phase relationship between the hind limb movement and hip bending in Fig. \ref{fig:coordination}a (right panel). The relationship between the fore limb movement and shoulder bending was shown in Fig.~S1. We observed a stronger in-phase relationship between the hind limb and hip bending.  We suspected that this is due to the relatively poor visibility of the fore limbs in the field-recorded videos, and the low magnitude of the shoulder bending compared to hip bending. 

The observations of the phase relationship between limb movements and body bending allow us to reduce the shape variables of short limbed, elongate lizard locomotion into two-dimensions. As discussed earlier, the body undulation profile $\kappa(s,t)$ can be approximated by a linear combination of $\sin(2\pi\xi s)$ and $\cos(2\pi\xi s)$ (under coefficients $w_1$ and $w_2$). We took $\xi=0.65$ from our previous analysis ($\lambda\approx 1.5$ for \textit{B. taylori} and \textit{B. kadwa} Fig. \ref{fig:analysis}b). We can then infer the limb contact states from the choice of reduced shape variables $w_1$ and $w_2$ such that the shoulder (hip) bending is in phase with the fore (hind) limb movement. The explicit shape space can be found in Fig. \ref{fig:coordination}b (left panel). 

We then used the geometric mechanics framework \cite{kelly1995geometric,wilczek1989geometric,rieser2019geometric} to compare the effectiveness of standing and traveling waves in these short limbed, elongate lizards. Geometric mechanics was originally developed to study thrust generation via self deformation at low Reynolds numbers \cite{purcell1977life,wilczek1989geometric}. Since the thrust is generated from properly coordinated self-deformation, we refer such thrust generation as self-propulsion. Recent work has shown that geometric mechanics can offer insights in biological locomotion when the frictional forces dominate over the inertial forces \cite{rieser2019geometric,chong2021coordination}. 

In the geometric mechanics framework, we seek to connect the self-deformation patterns with their resulting performance. Specifically, we analyze the net displacement in the world coordinate frame by studying the sequence of internal shape changes (in our case, of lizards). The space spanning internal shapes is then called shape space (Fig. \ref{fig:coordination}b). For simplicity, we only analyze the quasi-static regime\footnote{Quasi-static motion then implies that inertial forces are negligible compared to the ground reaction forces.} of lizard locomotion, where there is zero acceleration on the center of mass (CoM) in lizards. In this way, the velocities in shape space (shape velocity) and body velocities are then connected by a matrix called the \textit{connection vector field} (e.g., Fig. S3)~\cite{hatton2013geometric}. A gait, a periodic sequence of shape changes, can be represented as a closed-loop path in the shape space. In Fig. \ref{fig:coordination}b (right panel), we compared the standing and traveling wave body movements and their corresponding limb contact sequences. The net displacement of a gait can be approximated by a line integral of the vector field along the gait path~\cite{hatton2011geometric}. From Stokes' theorem, the line integral of a closed-loop path over a vector field can be visualized by a surface integral over the curl of the vector field (the height function, or often referred to as constraint curvature function \cite{hatton2013geometric}). The height function for the short limbed, elongate lizards was computed in Fig. \ref{fig:coordination}c (left panel). In summary, instead of laborious calculation, we can investigate the seeming complicated and diverse lizard wave dynamics with a help of a pre-computed diagram, and analyze locomotion performance by evaluating the surface integral.

The actual force model of these lizards in the field are unknown. We chose to approximate them using a model granular medium (poppy seeds) to numerically calculate the connection vector field~\cite{li2013terradynamics,chong2021coordination}. To bound the uncertainty in ground reaction forces, we used different force models (rate independent Coulomb friction and rate dependent viscous fluid, Fig. S2) and achieved similar conclusions as in Fig. \ref{fig:coordination}c. 
Further, in the derivation of the local connection vector field, we assumed that the magnitude of limb retraction is 17\% of the total thrust (body undulation and limb retraction), a value similar to the relative limb size. 

Two stripes emerged in the height function with an oblique intersection, which we interpreted as corresponding to the coordination for limb movements.
To better understand the meaning of the height function, we recomputed the height function for two hypothetical lizards: lizards with only fore limbs (Fig. \ref{fig:coordination}c: middle panel) and lizards with only hind limbs (Fig. \ref{fig:coordination}c: right panel). 
One of these stripes emerged in each height function for the hypothetical lizards, supporting our hypothesis that each stripe corresponds with the coordination of one pair of limbs.
From the structure of the height function, we inferred that an elliptical gait path with $\sigma \approx$  0.5 can lead to the greatest displacement, which was qualitatively the range of $\sigma$ measured from animal experiments (Fig. \ref{fig:analysis}b).

\subsection*{Body weight distribution}

From the above analysis, we noticed that the presence of limbs significantly affects the dynamics in body movements. The limbs would not only generate thrust, but also support the body weight in conjunction with the ventral surface. In fully limbed lizards, almost the entire body weight is supported by the limbs, whereas in limbless lizards the ventral surface supports the entire body weight. But for short limbed, elongate lizards, how should the body weight be distributed between the limbs and the ventral body surface for effective locomotion? We used geometric mechanics modeling to predict the optimal body weight distribution for short limbed, elongate lizards.

Quadrupedal locomotors typically utilize two types of limb contact patterns: the diagonal couplet and the lateral couplet~\cite{hildebrand1965symmetrical}. In the diagonal couplet, the limbs in the ground-contact phase are distributed in pairs along the diagonal (FR/HL) or counterdiagonal (FL/HR), where the body weight can be stably supported by the limbs (Fig. \ref{fig:distribution}a). In the lateral couplet, the limbs in the ground-contact phase are on the same side, which cannot stably support the entire body weight (Fig. \ref{fig:distribution}a). Thus, some ventral surface support is essential for the lateral couplet. We quantified the fraction of body weight supported by the limbs via $\gamma$. There is a limit on the force that the limbs can support without the animals tipping over (the torque between lateral couplets is greater than the torque from gravity) in the lateral couplet (Fig. \ref{fig:distribution}a). The detailed derivation to compute the body weight distribution can be found in~\cite{Baxi2021hildebrand}.

We compared three typical limb contact patterns: the pace, the lateral sequence (LS), and the trot, where leg phase shifts (fraction of a period that the hind limb leads the ipsilateral fore limb) were $0$, $0.25$, and $0.5$ respectively (Fig. \ref{fig:distribution}b). In the pace, the contact patterns were entirely lateral couplets; in the trot, the contact patterns were entirely diagonal couplets; in the LS, there was a mix of lateral and diagonal couplets. Assuming the duty factor to be 0.5, the fraction of the lateral couplet in the pace, the LS, and the trot were 1, 0.5, and 0 respectively. Further, the body bending spatial frequency $\xi$ was 1, 0.75, and 0.5 for the pace, LS, and trot, respectively, to enforce the in-phase relationship between the fore (hind) limb movements and shoulder (hip) bending.  

We conducted numerical simulations to predict the relationship between $\gamma$ and forward speed. We studied lizards with short limbs (a hypothetical locomotor with $l=0.05$, shorter limbs than \emph{B. muntingkamay}), lizards with intermediate limbs (a hypothetical locomotor with $l=0.17$, similar to \emph{B. kadwa} and \emph{B. taylori}), and lizards with long limbs (a hypothetical locomotor with $l=0.30$, similar to \emph{U. scoparia}). We observed that for short limbed lizards, it was optimal to use only body undulation to generate thrust ($\gamma = 0$, the pace, Fig. \ref{fig:distribution}c.1 ) while intermediate limbed lizards optimally used a hybrid thrust generation mechanism using both body undulation and limb retraction ($\gamma = 0.4$, the LS, Fig. \ref{fig:distribution}c.2). Finally, one available optimum for long limbed lizards is to solely use limbs to generate thrust ($\gamma = 0$, the trot, Fig. \ref{fig:distribution}c.3).

Thus, we showed that limbs are crucial to locomotion by short limbed, elongate lizards because they contribute to thrust as well as sharing some body weight with the ventral surface of the body, which can modulate lifting forces and thus thrust.
Depending on the limb size, our model suggests that lizards should properly distribute their body weight between the limbs and the ventral surface to generate effective locomotion. Therefore, we predicted that a traveling wave enhances locomotor performance as the body weight distribution (and thus thrust generation mechanism) shifts from the limbs to the body.

\begin{figure}
\centering
\includegraphics[width=1\linewidth,trim=0 0 0 0 ,clip]{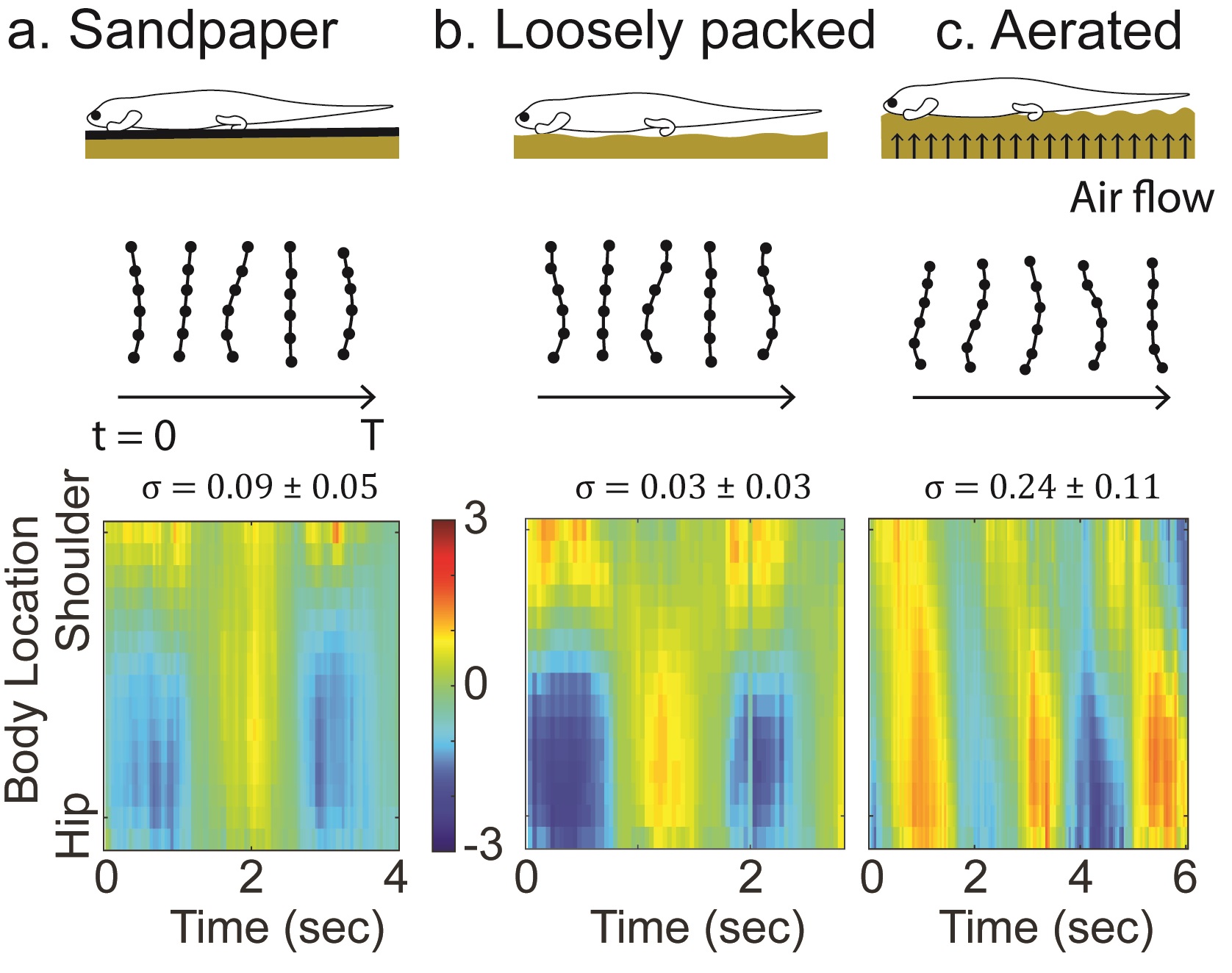}
\caption{\textbf{Traveling wave in fully limbed lizards induced by substrate variation} Comparison of the body wave dynamics of \textit{Uma scoparia} (a) on sandpaper and (b) on a loosely packed granular medium, and (c) on an aerated granular medium. An almost perfect standing wave is observed for \textit{Uma scoparia} on sandpaper and on the loosely packed granular medium, while features of a traveling wave emerge for \textit{Uma scoparia} on the aerated granular medium. Resulting $\sigma$ and $\lambda$ are shown in Fig. \ref{fig:analysis}b. The units of the colorbar are SVL$^{-1}$ for all panels. 
}
\label{fig:animalexp}
\end{figure}

\begin{figure}[!ht]
\centering
\includegraphics[width=0.9\linewidth,trim=0 0 0 0 ,clip]{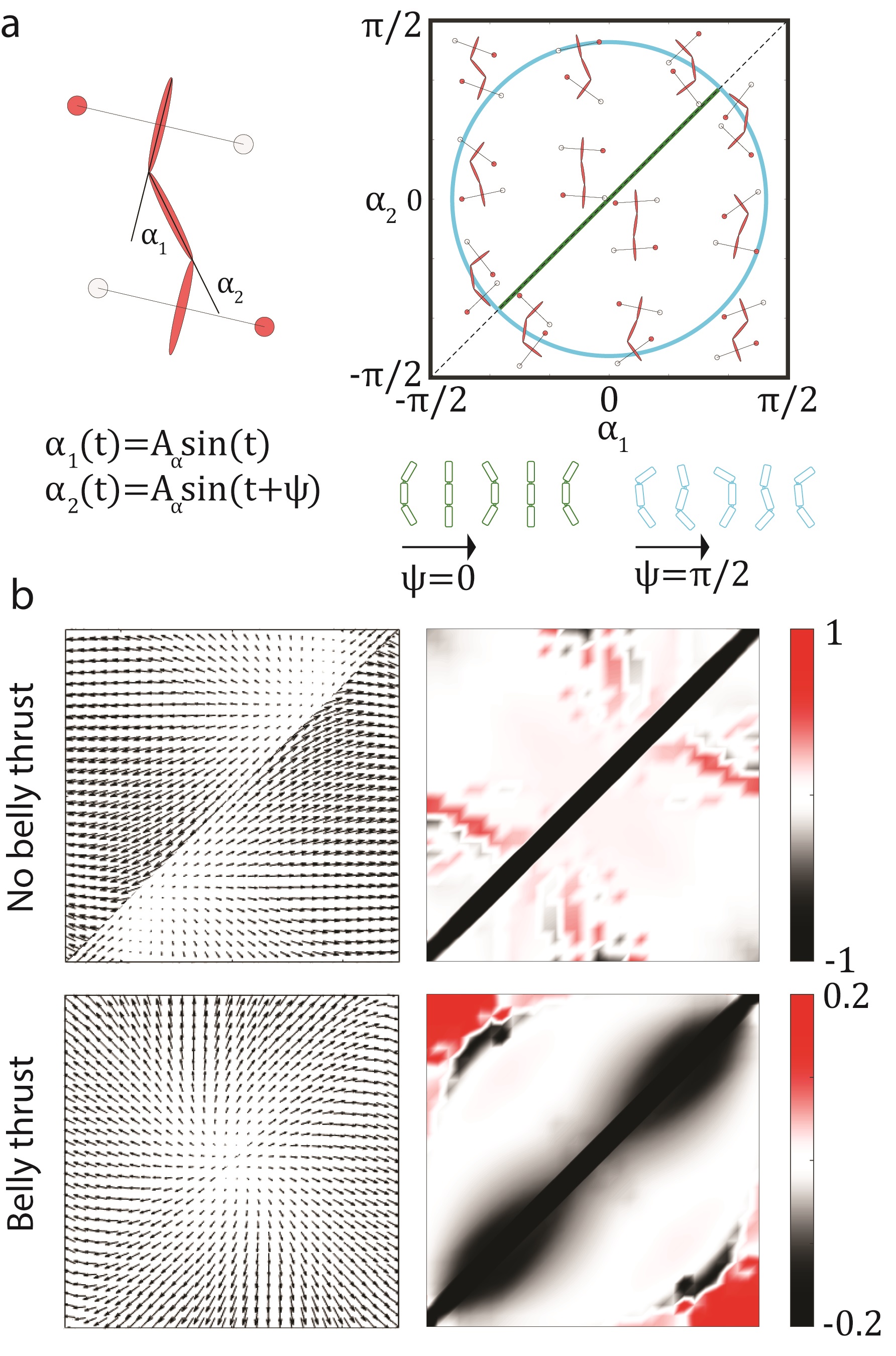}
\caption{\textbf{Geometric mechanics modeling for the robophysical experiments} (a) The definition of $\alpha_1$ and $\alpha_2$ and the body-limb coordination in the 3-link swimmer and 4 leg contacts. $\psi$ is the phase lag between the upper back and lower back actuators. Right panel demonstrates how leg contact patterns are coupled to the shape variables ($\alpha_1$ and $\alpha_2$). On the lower right half of the shape space, the contact patterns are counter-diagonal; on the upper left half of the shape space, the contact patterns are diagonal. Examples of the standing wave ($\psi=0$) and the traveling wave ($\psi = \pi/2$) are compared in the shape space. (b) Vector field and height functions for modelling the robophysical experiments on poppy seeds. The displacement can be approximated by the surface integral enclosed by the gait path over the height function (right panels). The units of the colorbar are ($10^{-3}\times \text{SVL}^{-1}/\text{rad}^{2}$). Units of axes in (b) are identical to the shape space in (a).}
\label{fig:robot}
\end{figure}

\begin{figure*}
\centering
\includegraphics[width=1\linewidth,trim=0 0 0 0 ,clip]{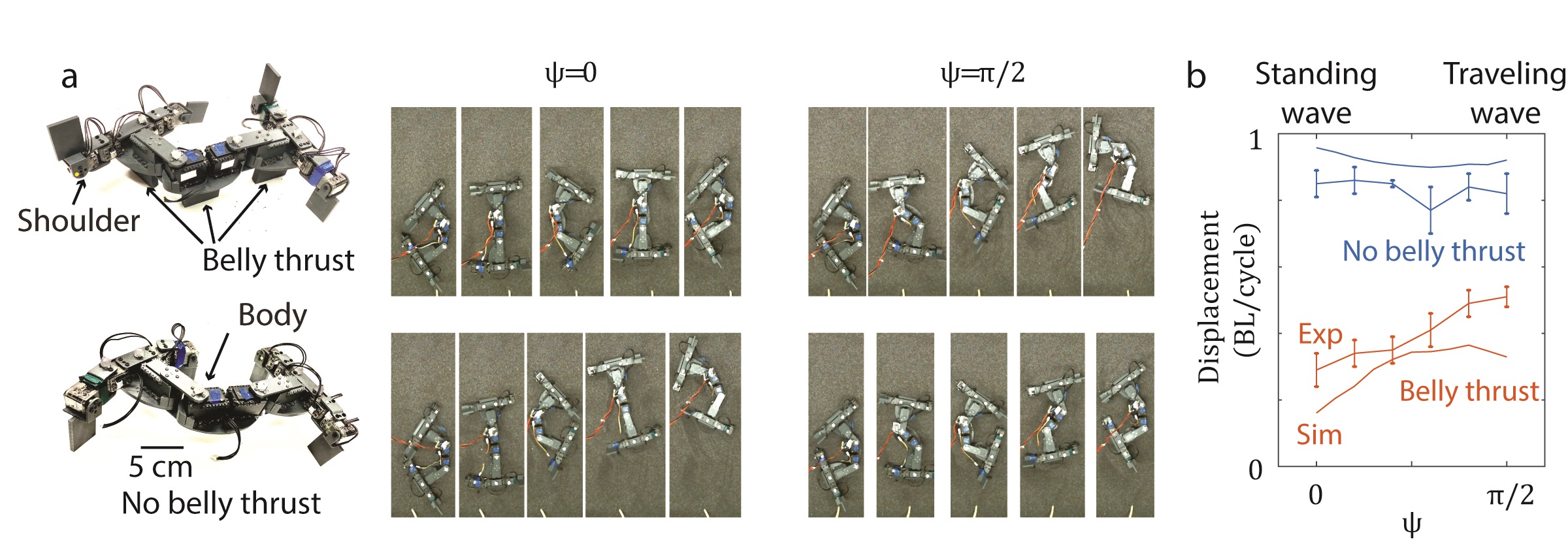}
\caption{\textbf{Robophysical experiments} (a) Snapshots of robots (top panel: belly thrust; bottom panel: no belly thrust) implementing standing wave ($\psi = 0$) and traveling wave ($\psi = \pi/2$) gaits. (b) The effect of $\psi$ on locomotion performance for robot with no belly thrust (blue curve) and robot with belly thrust (orange curve).}
\label{fig:robotexp}
\end{figure*}

\subsection*{Terrestrial swimming}

In fully limbed lizards, nearly the entire body weight is supported by the limbs. It is then commonly believed that at low speeds, lizards use standing wave body bending to coordinate with their limb movements~\cite{farley1997mechanics,ijspeert2007swimming,chong2021coordination}. 
Our geometric mechanics modeling predicts that the body weight distribution will affect how much a traveling wave contributes to thrust.
We tested this hypothesis by manipulating the substrate on which fully limbed lizards moved and investigated whether we could stimulate terrestrial swimming in fully limbed lizards.

To modulate the body weight distribution, we used an upward air flow through a granular medium to control the ground penetration resistance (GPR)\footnote{GPR is defined as the vertical ground resistance force per depth during intrusion}~\cite{cairns2011influence}, maintaining airflow below the onset of fluidization\footnote{Fluidization of granular media characterized by GPR dropping to zero} as in \cite{qian2015principles}. This technique proved useful in previous biological and robotics studies to evaluate locomotors' performance on flowable ground of various penetration resistance \cite{qian2015principles}. In doing so, the lifting forces at the limbs no longer fully supported the body weight and therefore some finite resistance lifting force acted on the ventral surface. When the fully limbed lizards (\textit{U. scoparia} and \textit{S. olivaceus}) ran across the region with reduced ground penetration force, they exhibited features of a traveling wave, indicated by the propagation of nodes (Fig. \ref{fig:animalexp}). We compared the wave flatness ($\sigma$) and wavelength $\lambda$ of the body undulations (Fig. \ref{fig:analysis}b) for lizards on the aerated granular medium, the loosely packed granular medium, and sandpaper. We found no difference in $\sigma$ between sandpaper and the loosely packed medium, but noted a significantly higher $\sigma$ on the aerated medium than the loosely packed medium for both species (\textit{U. scoparia}: $t=2.94$, $df=11$, $p=0.013$; \textit{S. olivaceus}: $t=2.43$, $df=9$, $p=0.038$), indicating a higher degree of traveling wave.

\subsection*{Robophysical experiments}

In the previous sections, we showed that although thrust generation in lizards results from a complex coordination of limb and body movements, we could modulate the degree to which traveling wave undulations were used by modulating the ground penetration resistance. We further explored the relative advantages of traveling waves and standing waves using a robophysical model where we can precisely control the thrust generation mechanism. Our robophysical model had four actuated limbs and two actuated body bending joints. The body shape of the robot can be uniquely described by the body joint angles: upper back $\alpha_1$ and lower back $\alpha_2$ (Fig. \ref{fig:robot}a, left panel).  Note that we chose two DoF body joints because two actuated body joints are minimal DoF that can enable a traveling wave (i.e., a 3-link swimmer). We designed removable belly intrusion plates to control the belly thrust generation mechanics. We compare the robot with belly thrust (top panel) and the robot without belly thrust (bottom panel) in Fig. \ref{fig:robotexp}a. The shoulder joints control the contact patterns of each limb. For simplicity, we only considered two combinations of contact patterns: diagonal contact and the counter-diagonal contact. 

As with the geometric mechanics models presented earlier, the gait of the robot could also be represented by a closed path in shape space (Fig. \ref{fig:robot}a). For simplicity, we considered the upper back and lower back as oscillating sinusoidal waves: $\alpha_1 (t) = A_{\alpha}\sin{(t)}$, $\alpha_2 (t) = A_{\alpha}\sin{(t+\psi)}$, where $A_{\alpha}$ is the amplitude, $\psi$ is the phase lag between the upper back and the lower back. A typical traveling wave can be described such that the upper back and lower back are $\pi/2$ out of phase \cite{hatton2013geometric}: $\psi=\pi/2$, which leads to a circular path in the shape space (blue curve in Fig. \ref{fig:robot}a). A typical standing wave can be described such that the upper back and lower back are in phase: $\psi=0$, which leads to a flattened ellipse (with eccentricity = 1) in the shape space (green curve in Fig.~\ref{fig:robot}a). 

We used contact pattern design algorithms to determine the coordination between the contact pattern and the body movements \cite{chongmoving}. The optimal coordination is shown in Fig. \ref{fig:robot}a's right panel, in agreement with our data on body-limb coordination in the biological experiments. We then tested the effect of $\psi$ on the robot\footnote{Note that some regions of shape space contain shapes where parts of the robot collide with other parts (e.g., upper right corner and lower right corner). The amplitude $A_\alpha$ was chosen such that the gait path does not pass through the self-collision region}. Snapshots of the robot implementing standing and traveling waves are shown in Fig. \ref{fig:robotexp}a. All experiments were conducted with at least 5 trials. The experimental results are shown in Fig. \ref{fig:robotexp}b. We found that traveling waves only increase speed (measured by distance moved per cycle) in the robot with belly thrust (Pearson's $\rho = 0.883$, $p=0.001$), whereas there are no significant differences between standing and traveling waves when robot lacks belly thrust (Pearson's $\rho = -0.147$, $p=0.438$). 

We used geometric mechanic height functions to explain our observations (Fig. \ref{fig:robot}b, right panel). As before, the displacement can be approximated by the surface integral over the gait path in the shape space. From the structure of the height function, we observed that the standing wave body bending can sufficiently coordinate with the robot without belly thrust since most of the negative volumes (indicated by black color) are distributed along the narrow diagonal line, which can be sufficiently bounded by a flattened ellipse. On the other hand, the traveling wave body undulation can better coordinate the robot with belly thrust because the negative volumes are distributed widely around the diagonal line. In the latter case, higher surface integral can be achieved for ellipses with increasing $\psi$ (and thus increasing $\sigma$). Further, the magnitude of the height function for locomotors without belly thrust (top panel in Fig. \ref{fig:robot}b) is much higher than that with belly thrust (bottom panel in Fig. \ref{fig:robot}b), indicating that a robot without belly thrust should have higher speed than one with belly thrust. 
The trends in the theoretical predictions and experimental data agreed, but we posit that the discrepancy in magnitude (Fig. \ref{fig:robotexp}b) was due to the accumulation of granular media \cite{mcinroe2016tail} as the robot implemented its gait. We interpreted our observations on traveling and standing waves by analyzing the connection vector field in Fig. \ref{fig:robot}b (left panel) \cite{bhatia2012helmholtz}. The connection vector field in locomotors with no belly thrust is almost curl-free (Fig. S3), which indicates that the contribution of body bending postures is almost path independent. In other words, only the body postures at the contact transitions matter. On the other hand, the connection vector field in locomotors with belly trust has non-negligible curls (Fig. S3), which indicates that the trajectory of body bending (standing wave vs. traveling wave) will affect the locomotor performance.

\section*{Discussion and Conclusion}
Lizards have evolved a diversity of body forms from fully limbed and short-bodied to limbless and elongate. We showed that this diversity in morphology coincides with a similar diversity in locomotion patterns, ranging from standing wave to traveling wave body undulation. We observed that the degree of body elongation and limb reduction were closely related to how the body and limb movements were coordinated, indicating an interconnected morphological and locomotor continuum. Using biological experiments, a geometric theory of locomotion, and robophysical experiments, we showed that the body weight distribution between the limbs and the body (and therefore, the primary thrust generation mechanism) plays a crucial role in the locomotor transition from fully limbed to limbless. Specifically, we found that fully limbed lizards adopted a traveling wave, something we termed terrestrial swimming when the penetration resistance of the substrate was reduced, resulting in belly contact. Furthermore, our robophysical experiments showed that a traveling wave only enhanced locomotor performance when some thrust was generated by the body.


One of the novel contributions presented in this paper was to use geometric mechanics as a tool to analyze seemingly complicated lizard locomotion. Specifically, we formulated different body wave dynamics as different paths in the shape space. The diagrammatic analysis by geometric mechanics then allowed us to visually and intuitively compare different wave dynamics. In this sense, our analysis simplified the laborious calculations, which would otherwise be required to study the diversity in lizard body wave dynamics. In addition, the geometric mechanics served as a bridge connecting the biological experiments and robophysical experiments, allowing us to systematically test gaits and conditions that are less commonly seen in biological systems. 

Limb reduction and body elongation result in a shift in body weight distribution from the limbs to the body \cite{renous1998respective}. We showed a traveling wave of body undulation enhanced locomotor performance during this transition. However, traveling wave body undulation requires larger local body curvatures, more complex neuromechanical control (to propagate the node in undulation)~\cite{ijspeert2007swimming}, and more degrees-of-freedom (at least two DoF) than standing wave undulation (one DoF). That a fully limbed lizard adopted terrestrial swimming when crossing a medium with low penetration resistance suggests that the degrees-of-freedom and neuromechanical control necessary for traveling waves may be widespread among lizards. Our work is the first to show that the coordination between body undulation and the limbs is a key feature of locomotion within the morphological transition between fully limbed, short bodied and limbless, elongate forms. We used \textit{Brachymeles} as morphologically intermediate species because they have similar levels of development of their fore and hind limbs~\cite{siler2011evidence,bergmann2019convergent}, but the limbed species of this genus are secondarily limbed, having re-evolved their limbs from a limbless ancestor~\cite{bergmann2020locomotion}. Therefore, our results should not be interpreted as representing an evolutionary transition in locomotion. Despite this, the geometric mechanics and robophysical approaches we used are naïve to evolutionary history, and our observations on \textit{Brachymeles} and the other, unrelated species that we used, closely coincide closely with these approaches, suggesting that biomechanics may dictate locomotor patterns in many of these convergent evolutions of snake-like forms. The role of how the evolutionary history affects locomotion of these forms could be further tested in a clade like \textit{Lerista}, which has evolved snake-like forms from limbed, short bodied ancestors~\cite{skinner2008rapid,morinaga2020evolution}. We also expect that our work on body and limb dynamics in these lizards will inform control of robots that need to traverse complex terrain.

\section*{Materials and Methods}

\subsection*{Tracking}

Positional data were extracted from videos with animal pose estimation software DeepLabCut (DLC) \cite{nath2019using}. Ten frames from each video were extracted and manually labeled. DLC would then provide positions for labeled points on all of the other frames. For each video, 21 points following the middle of the lizard body were labeled including 5 head points, 13 body points, and 3 tail points. If the species possessed limbs, 4 additional points were labeled marking each limb. 

\subsection*{Data Analysis}

To quantify the time-varying body shape of lizards, we obtained the body curvature from the tracked body positions. We used the methods introduced in \cite{nguyen2007curvature} to estimate the curvature from a relatively noisy backbone curve. 

Once we obtained the spatio-temporal information of the body curvatures, $\kappa(s,t)$, we reconstructed the body curvature profile by fitting with the least error. The error is defined as $||\kappa - B_\xi^T(B_\xi B_\xi^T)^{-1}B_\xi \kappa||_2$, where $B_\xi = [\cos(\xi s)\ \sin(\xi s)]$ is the shape basis function; and $\xi$ is the spatial frequency. The flatness of the trajectory in reduced shape space was approximated using PCA:  $\sigma = e(2)/e(1)$, where $e(i)$ is the percentage of the total variance explained by $i$-th principal component. The phases of shoulder (and hip) body bending and fore (and hind) limb movements were estimated using Hilbert Transforms.  

\subsection*{Animal experiments}

Experiments on stereotyped lizards, \textit{U. scoparia} ($N=7$) and \textit{S. olivaceus} ($N=6$), were performed in a 300 cm by 40 cm trackway filled with small glass spheres (mean $\pm$ s.d. diameter = 250 $\pm$ 30 $\mu$ m) to a depth of 20 cm. Experiments on snakes, \textit{C. occipitalis} ($N=11$), were performed in the laboratory on 300 $\mu$ m glass particles. 
Experiments on short limbed elongate lizards, \emph{B. kadwa} ($N=14$), \emph{B. taylori} ($N=8$) and, \emph{B. muntingkamay} ($N = 7$), were performed in the field on soil. We used soil from the habitats that they were found in, dried and sieved with a  \# 35 sieve, giving us 0.5-2mm diameter particles (midpoint 1.25mm), at 1 cm depth. \textit{L. praepedita} experiments were run in Australia on sieved dry sand that was 0.25-0.50 mm diameter, midpoint 0.375mm, at 1 cm depth. We chose the low depth because otherwise the animals would just bury themselves.

\subsection*{Robophysical Experiments}

The robophysical model (Fig. \ref{fig:robotexp}a) used in experiments is a 0.5 kg, quadrupedal robot with two body bending joints and four limbs. Body bending joints are capable of $\pm 90^\circ$ of rotation, actuated by Dymamixel AX-12A servo motors. Each limb possesses one degree of freedom: the shoulder joint to control the lifting and landing of the limb, actuated by a Dymamixel XL-320 servo motor. Limb paddles are 45 mm $\times$ 35 mm $\times$ 5 mm cuboids, and belly panels are 45 mm $\times$ 55 mm $\times$ 10 mm cuboids, all 3D-printed with ABS plastic. Two robot conditions are compared: a robot with belly thrust (robot with belly plate installed, Fig. \ref{fig:robotexp}a top) and a robot without belly thrust (robot with no belly plate Fig. \ref{fig:robotexp}a bottom).

We emulated the trackways used in animal experiments for the robot using a 2.1 m long, 0.5 m wide granular media bed trackway. We filled the trackway with poppy seeds of $\sim$1 mm diameter to a depth of 12 cm.
We had 4 leaf blowers (Toro 51599, 300 LPM) connected below the trackway that forced a continuous flow of air through a porous flow distributor to even the surface of the granular substrate every time before starting the experiment.
To track the robot's motion in its position space, 6 IR reflective markers were attached to the body of the robot. For tracking the trajectories of the markers, an OptiTrack motion capture system with 4 OptiTrack Flex 13 cameras were used to capture real-time 3D positions of the markers at 120 FPS frame rate. 
Displacement of the robot was then calculated from the tracked marker trajectory using MATLAB. 

\paragraph*{Acknowledgement}The authors are grateful for funding from NSF-Simons Southeast Center for Mathematics and Biology (Simons Foundation SFARI 594594), NSF grant IOS-1353703, President's Undergraduate Research Awards at Georgia Institute of Technology, and Army Research Office Grant W911NF-11-1-0514. We thank Prof. Robert J. Full for helpful discussion on laboratory lizard experiments. We thank Andras Karsai and Christopher Pierce for proofreading.

\newpage

\bibliographystyle{plainnat}
\bibliography{references}

\end{document}